# The expansion of orthogonal complete set and teleportation of an arbitrary two-qubit state

Xin-Wei Zha [*] and Cun-Bing Huang

(Department of Applied Mathematics and Physics, Xi'An Institute of Post and Telecommunications,

Xi'an, People's Republic of China 710061)

In accordance with the principle of superposition and operator rule, the state of the whole system composed of the state of the particles to be teleported and quantum channel can be expanded by Bell bases and transformation operator. Theoretically, if determinant of transformation operators is not zero, the teleportation can be realized only by performing an inverse transformation. Actually, if the transformation operator is not a unitary operation, then by using one auxiliary qubits, the teleportation can be realized only by performing a collective unitary transformation. Moreover, the further analysis of the relationship between collective unitary operation and transformation operators is discussed.

[*]Correspondi
[*]Corresponding Author: Tel: +86-29-85392889
  Email:   Zhxw@xiyou.edu.cn (Zha Xin-Wei)

# 1. INTRODUCTION

Quantum teleportation has received much attention both theoretically and experimentally in recent years due to its important applications in quantum calculation and quantum communication. Quantum teleportation was originally proposed by Bennett *et al.* [1]. Many schemes of quantum teleportation were presented [2,3]. Recently, Yan Fengli and Ding Hewei propose a scheme for probabilistic teleportation of unknown two-particle state with partly entangled four-particle state via POVM [4]. It will be shown that by performing two Bell state measurements, a proper POVM and a unitary transformation, the unknown two-particle state can be teleported from the sender Alice to the receiver Bob with certain probability. In this scheme, they need to use two auxiliary qubits. However, all of the proposed schemes do not have any systematical analysis.

In this paper, in accordance with the principle of superposition, the system state of particles can be expanded by Bell bases and transformation operator. Through discussing that if the transformation operator is reversible, quantum teleportation can be realized, and at the same time if the transformation operator is unitary, quantum teleportation can successfully act while Bob only need to perform a reverse transformation. If the transformation operator is reversible and not unitary, Bob only need to use one auxiliary qubit.

## 2. BELL BASES EXPANSION AND TRANSFORMATION OPERATION IN TELEPORTATION.

As ref. [4] suppose that the sender Alice has two particles 1, 2 in an unknown state

$$|\chi\rangle_{12} = (a|00\rangle + b|01\rangle + c|10\rangle + d|11\rangle)_{12} \qquad (1)$$

Where a, b, c, d are arbitrary complex numbers, and wave function satisfy normalization conditions $|a|^2 + |b|^2 + |c|^2 + |d|^2 = 1$. That Alice and Bob share quantum channel in the form of following partly pure entangled four-particle state,

$$|\varphi\rangle_{3456} = (\alpha|0000\rangle + \beta|1001\rangle + \gamma|0110\rangle + \delta|1111\rangle)_{3456} \qquad (2)$$

The particles particle pair (1,2) and 3 and 4, are in Alice's possession, and other two particles 5 and 6 are in Bob's possession. The system state of six particles is

$$|\phi\rangle_w = |\chi\rangle_{12} \otimes |\varphi\rangle_{3456} \qquad (3)$$

In accordance with the principle of superposition and transformation operator [5,6], the wave function of $|\phi\rangle_w$ can be represented in the form of a series

$$|\psi\rangle_{123456} = |\chi\rangle_{12} \otimes |\varphi\rangle_{3456} = \frac{1}{4}\sum_{i=1}^{4}\sum_{j=1}^{4} \varphi_{14}^{i}\varphi_{23}^{j}\sigma_{56}^{ij}|\chi\rangle_{56} \qquad (4)$$

Where $\varphi_{14}^{i}, \varphi_{23}^{j}$ are the Bell states, For example,

$$\varphi_{14}^{1} = \frac{1}{\sqrt{2}}(|00\rangle + |11\rangle)_{14},$$

$$\varphi_{14}^{2} = \frac{1}{\sqrt{2}}(|00\rangle - |11\rangle)_{14},$$

$$\varphi_{14}^{3} = \frac{1}{\sqrt{2}}(|01\rangle + |10\rangle)_{14} \qquad (5)$$

$$\varphi_{14}^{4} = \frac{1}{\sqrt{2}}(|01\rangle + |10\rangle)_{14}$$

$$|\chi\rangle_{56} = (a|00\rangle + b|01\rangle + c|10\rangle + d|11\rangle)_{56} \qquad (6)$$

$\sigma_{56}^{ij}$ is called the transformation operator.

In order to realize the teleportation, firstly Alice performs two Bell state measurements on particles 2,3 and 1,4, then the resulting state of Bob's particles 5,6 will be states $|\psi\rangle_{56}^{ij} = \frac{1}{4}\sigma_{56}^{ij}|\varphi\rangle_{56}$, For example,

$$|\psi\rangle_{56}^{11} = {}_{14}\langle\varphi^{1}|{}_{23}\langle\varphi^{1}|\varphi\rangle_{w} = \frac{1}{4}\sigma_{56}^{11}|\varphi\rangle_{56} \qquad (7)$$

By operator transformation [5,6], It can be shown

$$\sigma_{56}^{11} = 2\begin{pmatrix} \alpha & 0 & 0 & 0 \\ 0 & \beta & 0 & 0 \\ 0 & 0 & \gamma & 0 \\ 0 & 0 & 0 & \delta \end{pmatrix} \tag{8}$$

The four product basis of two-fold tensor products are chosen to be

$$|00\rangle = \begin{pmatrix} 1 \\ 0 \\ 0 \\ 0 \end{pmatrix}, \quad |01\rangle = \begin{pmatrix} 0 \\ 1 \\ 0 \\ 0 \end{pmatrix}, |10\rangle = \begin{pmatrix} 0 \\ 0 \\ 1 \\ 0 \end{pmatrix}, |11\rangle = \begin{pmatrix} 0 \\ 0 \\ 0 \\ 1 \end{pmatrix} \tag{9}$$

So,

$$\begin{aligned}\sigma_{56}^{11}|\varphi\rangle_{56} &= \sigma_{56}^{11}(a|00\rangle + b|01\rangle + c|10\rangle + d|11\rangle)_{56} \\ &= 2(\alpha a|00\rangle + \beta b|01\rangle + \gamma c|10\rangle + \delta d|11\rangle)\end{aligned} \tag{10}$$

Similarly, it can be obtained:

$$\sigma_{56}^{12} = 2\begin{pmatrix} \alpha & 0 & 0 & 0 \\ 0 & -\beta & 0 & 0 \\ 0 & 0 & \gamma & 0 \\ 0 & 0 & 0 & -\delta \end{pmatrix} \quad \sigma_{56}^{13} = 2\begin{pmatrix} 0 & \alpha & 0 & 0 \\ \beta & 0 & 0 & 0 \\ 0 & 0 & 0 & \delta \\ 0 & 0 & \gamma & 0 \end{pmatrix}$$

$$\sigma_{56}^{14} = 2\begin{pmatrix} 0 & -\alpha & 0 & 0 \\ \beta & 0 & 0 & 0 \\ 0 & 0 & 0 & \delta \\ 0 & 0 & -\gamma & 0 \end{pmatrix} \quad \sigma_{56}^{21} = 2\begin{pmatrix} \alpha & 0 & 0 & 0 \\ 0 & \beta & 0 & 0 \\ 0 & 0 & -\gamma & 0 \\ 0 & 0 & 0 & -\delta \end{pmatrix}$$

$$\sigma_{56}^{22} = 2\begin{pmatrix} \alpha & 0 & 0 & 0 \\ 0 & -\beta & 0 & 0 \\ 0 & 0 & -\gamma & 0 \\ 0 & 0 & 0 & \delta \end{pmatrix} \quad \sigma_{56}^{23} = 2\begin{pmatrix} 0 & \alpha & 0 & 0 \\ \beta & 0 & 0 & 0 \\ 0 & 0 & 0 & -\delta \\ 0 & 0 & -\gamma & 0 \end{pmatrix}$$

$$\sigma_{56}^{24} = 2\begin{pmatrix} 0 & -\alpha & 0 & 0 \\ \beta & 0 & 0 & 0 \\ 0 & 0 & 0 & -\delta \\ 0 & 0 & \gamma & 0 \end{pmatrix} \quad \sigma_{56}^{31} = 2\begin{pmatrix} 0 & 0 & \alpha & 0 \\ 0 & 0 & 0 & \beta \\ \gamma & 0 & 0 & 0 \\ 0 & \delta & 0 & 0 \end{pmatrix} \quad (11)$$

$$\sigma_{56}^{32} = 2\begin{pmatrix} 0 & 0 & -\alpha & 0 \\ 0 & 0 & 0 & -\beta \\ \gamma & 0 & 0 & 0 \\ 0 & \delta & 0 & 0 \end{pmatrix} \quad \sigma_{56}^{33} = 2\begin{pmatrix} 0 & 0 & 0 & \alpha \\ 0 & 0 & \beta & 0 \\ 0 & \gamma & 0 & 0 \\ \delta & 0 & 0 & 0 \end{pmatrix}$$

$$\sigma_{56}^{34} = 2\begin{pmatrix} 0 & 0 & 0 & -\alpha \\ 0 & 0 & \beta & 0 \\ 0 & -\gamma & 0 & 0 \\ \delta & 0 & 0 & 0 \end{pmatrix} \quad \sigma_{56}^{41} = 2\begin{pmatrix} 0 & 0 & -\alpha & 0 \\ 0 & 0 & 0 & -\beta \\ \gamma & 0 & 0 & 0 \\ 0 & \delta & 0 & 0 \end{pmatrix}$$

$$\sigma_{56}^{42} = 2\begin{pmatrix} 0 & 0 & -\alpha & 0 \\ 0 & 0 & 0 & \beta \\ \gamma & 0 & 0 & 0 \\ 0 & -\delta & 0 & 0 \end{pmatrix} \quad \sigma_{56}^{43} = 2\begin{pmatrix} 0 & 0 & 0 & -\alpha \\ 0 & 0 & -\beta & 0 \\ 0 & \gamma & 0 & 0 \\ \delta & 0 & 0 & 0 \end{pmatrix}$$

$$\sigma_{56}^{44} = 2\begin{pmatrix} 0 & 0 & 0 & \alpha \\ 0 & 0 & -\beta & 0 \\ 0 & -\gamma & 0 & 0 \\ \delta & 0 & 0 & 0 \end{pmatrix}$$

Obviously, the determinant of transformation operators $\sigma_{56}^{ij}$

$$\left|\det \sigma_{56}^{ij}\right| = 2\alpha\beta\gamma\delta \neq 0. \quad (12)$$

So, the transformation operators $\sigma_{56}^{ij}$ have reverse operator. Then Alice informs Bob two Bell state measurement outcomes via a classical channel. By outcomes received, Bob can determine the state of particles 5,6 exactly by transformation operators $(\sigma_{56}^{ij})^{-1}$.

## 3. RESULTS AND DISCUSSION.

### 3.1 Comparison with the results of Yan Fengli and Ding Hewei[4]

From above result we known, theoretically, if determinant of transformation operators is not zero, then Bob can determine the state of particles 5,6 exactly by transformation operators $(\sigma_{56}^{ij})^{-1}$. However, actually, the operator $(\sigma_{56}^{ij})^{-1}$ is not a unitary operation. In ref [4], In order to realize the teleportation, Bob introduces two auxiliary qubits a, b in the state $|00\rangle_{ab}$, Then Bob performs two controlled-not operations with particles 5,6 as the controlled qubits and the auxiliary particles a, b as the target qubits respectively.

It is easy to show:

$$\begin{aligned}
(CN)_{5a}(CN)_{6b}\sigma_{56}^{11} &= I_5 I_6 (\alpha I_a I_b + \beta I_a \sigma_{bx} + \gamma \sigma_{ax} I_b + \delta \sigma_{ax} \sigma_{bx}) \\
&= I_5 \sigma_{6z}(\alpha I_a I_b - \beta I_a \sigma_{bx} + \gamma \sigma_{ax} I_b - \delta \sigma_{ax} \sigma_{bx}) \\
&+ \sigma_{5z} I_6 (\alpha I_a I_b + \beta I_a \sigma_{bx} - \gamma \sigma_{ax} I_b - \delta \sigma_{ax} \sigma_{bx}) \\
&+ \sigma_{5z} \sigma_{6z}(\alpha I_a I_b - \beta I_a \sigma_{bx} - \gamma \sigma_{ax} I_b + \delta \sigma_{ax} \sigma_{bx})
\end{aligned} \quad (13)$$

Here $(CN)_{ij}$ is controlled-not operation

$$(CN)_{ij} = \begin{bmatrix} 1 & 0 & 0 & 0 \\ 0 & 1 & 0 & 0 \\ 0 & 0 & 0 & 1 \\ 0 & 0 & 1 & 0 \end{bmatrix}_{ij} = \begin{pmatrix} 1 & 0 \\ 0 & 0 \end{pmatrix}_i \otimes \begin{pmatrix} 1 & 0 \\ 0 & 1 \end{pmatrix}_j + \begin{pmatrix} 0 & 0 \\ 0 & 1 \end{pmatrix}_i \otimes \begin{pmatrix} 0 & 1 \\ 1 & 0 \end{pmatrix}_j \quad (14)$$

$$(CN)_{5a}(CN)_{6b}|\psi\rangle_{56}^{11}|00\rangle_{ab} = \frac{1}{4}(CN)_{5a}(CN)_{6b}\sigma_{56}^{11}|\varphi\rangle_{56}|00\rangle_{ab}$$

$$= \frac{1}{4}[(a|00\rangle + b|01\rangle + c|10\rangle + d|11\rangle)_{56} \otimes (\alpha|00\rangle + \beta|01\rangle + \gamma|10\rangle + \delta|11\rangle)_{ab}$$
$$+ (a|00\rangle - b|01\rangle + c|10\rangle - d|11\rangle)_{56} \otimes (\alpha|00\rangle - \beta|01\rangle + \gamma|10\rangle - \delta|11\rangle)_{ab} \quad (15)$$
$$+ (a|00\rangle + b|01\rangle - c|10\rangle - d|11\rangle)_{56} \otimes (\alpha|00\rangle + \beta|01\rangle - \gamma|10\rangle - \delta|11\rangle)_{ab}$$
$$+ (a|00\rangle + b|01\rangle - c|10\rangle + d|11\rangle)_{56} \otimes (\alpha|00\rangle - \beta|01\rangle - \gamma|10\rangle + \delta|11\rangle)_{ab}]$$

And Obviously, This express is the same as ref [4], therefore, as ref [4], At this stage, Bob makes an optimal POVM on the ancillary particle a, b to conclusively distinguish the above states.

### 3.2 Relationship between collective unitary operator and transformation operator

From e.q (11), we can obtain

$$\sigma_{56}^{12} = 2\begin{pmatrix} \alpha & 0 & 0 & 0 \\ 0 & -\beta & 0 & 0 \\ 0 & 0 & \gamma & 0 \\ 0 & 0 & 0 & -\delta \end{pmatrix} = 2\sigma_{56}^{11}(I_5 \otimes \sigma_{6z}) = 2(I_5 \otimes \sigma_{6z})\sigma_{56}^{11}$$

$$\sigma_{56}^{13} = 2\begin{pmatrix} 0 & \alpha & 0 & 0 \\ \beta & 0 & 0 & 0 \\ 0 & 0 & 0 & \delta \\ 0 & 0 & \gamma & 0 \end{pmatrix} = 2\sigma_{56}^{11}(I_5 \otimes \sigma_{6x}) = 2(I_5 \otimes \sigma_{6x})\begin{pmatrix} \beta & 0 & 0 & 0 \\ 0 & \alpha & 0 & 0 \\ 0 & 0 & \gamma & 0 \\ 0 & 0 & 0 & \delta \end{pmatrix} \quad (16)$$

And so on.

Therefore, the transformation operators $\sigma_{56}^{ij}$ can be expressed as

$$\sigma_{56}^{ij} = (\sigma_5^i \otimes \sigma_6^j)A_{56}^{ij} \quad (17)$$

And $A_{56}^{ij}$ are $4\times 4$ diagonal matrixes.

Therefore, after Bell state measurement, Alice informs Bob of her measured results via a classical channel. Next, Bob will try to reconstruct the original state with his particles 5 and 6. First, Bob needs to perform an unitary transformation $U_1^{ij}=\sigma_5^i\otimes\sigma_6^j$ on particles 5, 6,. Table I shows different measurement results by Alice and Bob's relevant of unitary transformations.

TABLE I. Unitary transformations $U_1^{ij}$ corresponding to the states of particles 5 and 6.

| Alice's measurement results of particles 14 and 23, respectively. | Bob's unitary transformations of particles 5 and 6. |
|---|---|
| $\varphi^+$ | I |
| $\varphi^-$ | $\sigma_z=\begin{pmatrix}1 & 0\\ 0 & -1\end{pmatrix}$ |
| $\psi^+$ | $\sigma_x=\begin{pmatrix}0 & 1\\ 1 & 0\end{pmatrix}$ |
| $\psi^-$ | $\sigma_y=\begin{pmatrix}0 & -1\\ 1 & 0\end{pmatrix}$ |

For instance, if the measurement result of Alice is $|\psi^+\rangle_{14}|\varphi^+\rangle_{23}$, then Bob performs unitary transformation $U_1^{31}$ on particles 5, 6, and $U_1=\sigma_5^3\otimes\sigma_6^1=\sigma_{5x}\otimes I_6$.

Furthermore, as ref [7], Bob introduces an auxiliary two-state particle $a$ with the initial state

$|0\rangle_a$ and performs a collective unitary transformation on particles 5, 6 and $a$. In order for Bob to reconstruct the original state under the basis $\{|\varphi_0\rangle_{56}|0\rangle_a, |\varphi_0\rangle_{56}|1\rangle_a\}$ (where $|\varphi_0\rangle_{56}$ stands for the basis of an four-dimensional Hilbert space), a unitary transformation ( a $8\times 8$ matrix) may take the form

$$U_2 = \begin{pmatrix} A_1 & A_2 \\ A_2 & -A_1 \end{pmatrix} \tag{18}$$

Where $A_1$ and $A_2$ are $4\times 4$ matrixes and can be expressed as

$$A_1 = diag(a_1, a_2, a_3, a_4) \tag{19}$$

And

$$A_2 = diag\left(\sqrt{1-a_1^2}, \sqrt{1-a_2^2}, \sqrt{1-a_3^2}, \sqrt{1-a_4^2}\right) \tag{20}$$

It is easy to show:

$$U_2|\psi\rangle_{56}|0\rangle_a = A_1|\psi\rangle_{56}|0\rangle_a + A_2|\psi\rangle_{56}|1\rangle_a \tag{21}$$

Therefore, if $(A_{56}^{ij})^{-1} = k\hat{A}_1$, k is a constant number, Then Bob measures the state of particle $a$.

If the measured result is $|1\rangle_a$, the teleportation is failed. If the measurement result is $|0\rangle_a$, the state of particles 5 and 6 collapses to

$$|\chi\rangle_{56} = (a|00\rangle + b|01\rangle + c|10\rangle + d|11\rangle)_{56} \qquad (22)$$

The teleportation is successfully realized.

4. **CONCLUSION**

In summary, we have found that the transformation operator is in related to quantum channel and Bell state measurement, however it have no connection with teleport quantum state. If the transformation operator is reversible, then teleportation can easily realized. As a result, the reversibility of the transformation operators is the necessary condition of teleportation.

5. **REFERENCES**

[1] Bennett C.H. et al., 1993 Phys. Rev. Lett. 70 1895.

[2] W. L. Li *et al.*, 2000 Phys. Rev. A **61**, 034301 .

[3] Lu.H. and Guo G.C., 2000. Phys. Lett. A 276  209.

[4] Yan F.L.and, Ding H.W. 2006. Chin. Phys. Lett. 23  17.

[5] Zha Xin-Wei 2002 Acta Physica Sinaca,51(4):723~726(in Chinese).

[6]Zha Xin-wei Zhang Chun-min 2006 Journal of Xi'an Jiaotong University, **40**(2) ):243~246(in Chinese).

[7] Fang J X, Lin Y S and Zhu S Q *et al* . 2003 *Phys Rev A,***67**(1)014305.

**ACKNOWLEDGEMENTS:**

This work is supported by Shaanxi Natural Science Foundation under Contract Nos. 2004A15 and Science Plan Foundation of office the Education Department of Shaanxi Province Contract Nos. 05JK288.